\documentclass[11pt]{article}
\usepackage{gtml2025_workshop}
\usepackage{amsmath} 
\usepackage{amsthm} 
\usepackage{amssymb} 
\usepackage{amssymb} 
\usepackage{url} %
\usepackage{verbatim} 
\usepackage{todonotes} 
\usepackage{aliascnt}
\usepackage{autonum}
\usepackage{xcolor}
\definecolor{darkblue}{rgb}{0.0,0.0,0.55}
\usepackage[colorlinks=true, allcolors=darkblue]{hyperref}
\usepackage{comment}
\usepackage{orcidlink}
\usepackage{subcaption}

\newaliascnt{lemma}{theorem}

\aliascntresetthe{lemma}

\newaliascnt{proposition}{theorem}

\aliascntresetthe{proposition}

\newaliascnt{corollary}{theorem}

\aliascntresetthe{corollary}

\newaliascnt{conjecture}{theorem}

\aliascntresetthe{conjecture}

\newaliascnt{assumption}{theorem}

\aliascntresetthe{assumption}

\newaliascnt{definition}{theorem}
\newtheorem{definition}[definition]{Definition}
\aliascntresetthe{definition}

\newaliascnt{question}{theorem}

\aliascntresetthe{question}

\newaliascnt{remark}{theorem}

\aliascntresetthe{remark}

\newaliascnt{example}{theorem}
\newtheorem{example}[example]{Example}
\aliascntresetthe{example}

\newaliascnt{postulat}{theorem}
\newtheorem{postulat}[postulat]{Postulate}
\aliascntresetthe{postulat}

\newtheorem*{notation*}{Notation}
\newtheorem*{theorem*}{Theorem}
\newtheorem*{conjecture*}{Conjecture}

\usepackage[capitalise,nameinlink]{cleveref}
\crefname{theorem}{theorem}{theorems}
\Crefname{theorem}{Theorem}{Theorems}
\crefname{lemma}{lemma}{lemmas}
\Crefname{lemma}{Lemma}{Lemmas}
\crefname{proposition}{proposition}{propositions}
\Crefname{proposition}{Proposition}{Propositions}
\crefname{corollary}{corollary}{corollaries}
\Crefname{corollary}{Corollary}{Corollaries}
\crefname{conjecture}{conjecture}{conjectures}
\Crefname{conjecture}{Conjecture}{Conjectures}
\crefname{assumption}{assumption}{assumptions}
\Crefname{assumption}{Assumption}{Assumptions}
\crefname{definition}{definition}{definitions}
\Crefname{definition}{Definition}{Definitions}
\crefname{question}{question}{questions}
\Crefname{question}{Question}{Questions}
\crefname{remark}{remark}{remarks}
\Crefname{remark}{Remark}{Remarks}
\crefname{example}{example}{examples}
\Crefname{example}{Example}{Examples}
\crefname{postulat}{postulate}{postulates}
\Crefname{postulat}{Postulate}{Postulates}

\title{Mathematical Foundations of \\Modeling ETL Process Chains}

\author{\textbf{L. Maier}$^{1}$, 
\textbf{L. Schulze}$^{2}$\,\orcidlink{0000-0003-0662-3279},
\textbf{R. Lilow}$^{1}$\,\orcidlink{0000-0002-6960-6065}, 
\textbf{L. Hahn}$^{1}$,
\textbf{N. Krasowski}$^{1}$\,\orcidlink{0009-0007-1680-9233},
\textbf{A. Barth$^{1}$\,\orcidlink{0000-0002-4117-7924}}, \\
\textbf{S. Gaebel}$^{1}$,
\textbf{F. Güran}$^{1}$,
\textbf{O. Hanau}$^{1}$,
\textbf{G. Wagner}$^{1}$,
\textbf{F. Borgmann}$^{1}$,
\textbf{O. Arenz}$^{2}$\,\orcidlink{0000-0002-9470-2833}, \\
\textbf{J. Peters}$^{2,3,4,5}\,\orcidlink{0000-0002-5266-8091}$ \\[6pt]
$^{1}$Deepshore GmbH, Germany \\
$^{2}$TU Darmstadt, Germany \\
$^{3}$German Research Center for AI (DFKI), Germany \\
$^{4}$Hessian.AI, Germany \\
$^{5}$Centre of Cognitive Science, Germany \\[4pt]
\texttt{info@deepshore.de, lucas.schulze@tu-darmstadt.de, oleg.arenz@tu-darmstadt.de}
}

\date{}

\begin{document}

\maketitle

\begin{abstract}
Extract-Transform-Load (ETL) processes are core components of modern data processing infrastructures. The throughput of processed data records can be adjusted by changing the amount of allocated resources, i.e.~the number of parallel processing threads for each of the three ETL phases, but also depends on stochastic variations in the per-record processing times. In chains of multiple consecutive ETL processes, the relation between allocated resources and overall throughput is further complicated, for example by the occurrence of bottlenecks affecting all subsequent ETL processes.
\\[0.3em]
We develop a mathematical model of ETL process chains that is accurate at the level of time-aggregated throughput and suitable for efficient simulation. The process chain is represented as a controlled discrete-time Markov process on a directed acyclic graph whose edges are individual ETL processes. We model the mean throughput as a bounded, monotone function of the number of parallel threads, to capture the diminishing benefit of allocating more threads. We furthermore introduce a Flow Balance postulate linking number of threads, mean throughput, and mean processing time. The stochastic processing times are then modeled by non-negative heavy-tailed distributions around the mean processing time.
\\[0.3em]
This framework provides a principled simulator for ETL networks and a foundation for learning- and control-based resource allocation.
\end{abstract}

\section{Introduction}
In our data-driven world, most services we take for granted rely on massive volumes of data being transferred, processed and analyzed every day---bank transactions, public transport, electronic payments, social media, to name just a few. Although the specific ways of handling the data are as diverse as the services themselves, they generally follow the same universal pattern of an ETL process: \textbf{E}xtract data from a source system, \textbf{T}ransform it, and finally \textbf{L}oad it into a target system. Often, multiple ETL processes are performed consecutively, to achieve more complex data handling tasks. To increase processing throughput, each of the three phases in an ETL process can handle multiple data records in parallel. However, the degree of parallelization determines the required allocation of computing resources to the individual phases. Due to the ever growing volume of data being processed, striking the balance between performance and resource efficiency becomes an increasingly important challenge. This requires tuning the resource allocation to multiple consecutive processing steps in a way that avoids bottlenecks and idling resources. Up to now, this is mainly achieved by human expertise, which is, however, labor-intensive and suboptimal, in particular if the number of incoming records varies over time. Although there are existing auto-scaling mechanisms assisting with this, these do not account for the dependencies between consecutive steps and also require human expertise for their configuration. Furthermore, any real deployment of ETL processes is inevitably subject to fluctuations in the processing performance caused by interactions with other IT processes running on the same hardware or independently accessing the same data stores. Since these environmental interactions are out of our control, it is crucial to adapt to such fluctuations by dynamically optimizing the resource allocation, for example, to maximize throughput while minimizing resource usage.

Such optimization could be performed using online or offline methods. Online methods would optimize resource allocation during operation by leveraging a mathematical model of the ETL process to predict future system behavior. A prominent example is Model Predictive Control (MPC)~\citep{rawlings2020model}, which optimizes actions (resource allocations) across a finite, receding time horizon. In this approach, the optimizer generates a sequence of allocations for several upcoming discrete time steps, but the controller applies only the action for the immediate next step. At each subsequent time interval, the optimization is repeated with a shifted horizon and an updated system state. Similarly, planning methods like Monte-Carlo Tree Search (MCTS)~\citep{bandit_monte_carlo} can be applied by iteratively building a tree of action sequences using random roll-outs from the model to find the best next action. Offline optimization, on the other hand, would be performed prior to deployment. For example, Reinforcement Learning (RL)~\citep{SuttonBarto2018, rl_resource_allocation} would learn a policy that maps the current system state to an action using the system model for simulation. By not relying on real-time optimization, such an approach can potentially produce better behavior by allowing for more extensive exploration of the state space during the training phase.

The aim of this work is to construct a mathematical model suitable for future applications of the aforementioned optimization approaches. To achieve this, we develop a novel modeling framework that enables realistic simulation and admits a formulation in the language of sequential decision making.

\textbf{The main contributions} of this work are as follows:
\begin{itemize}
    \item \textbf{Flow Balance postulate for ETL phases.}
    We introduce the Flow Balance postulate (\Cref{post:flow_balance}), linking parallel resource allocation, mean throughput, and mean processing time at the level of individual ETL phases. To the best of our knowledge, this relation has not been formalized previously in the context of ETL systems.

    \item \textbf{Bounded monotone throughput model.}
    We propose a mathematically convenient and empirically motivated class of bounded, monotone throughput curves (\Cref{post: space of convenient curves}), which captures diminishing returns of parallelism while remaining suitable for gradient-based optimization and calibration against data.

    \item \textbf{Markovian formulation of ETL dynamics.}
    We model the full ETL process chain as a controlled Markov process with explicit state, action, and transition definitions (\Cref{sec:DES}). This formulation captures bottlenecks and other interactions between linked ETL processes, enabling realistic simulation of complex ETL workflows.

    \item \textbf{A simulation model designed for learning and control.}
    By combining the above modeling choices, we obtain a discrete-event simulation that is both faithful to real ETL system behavior and explicitly compatible with reinforcement learning and other sequential decision-making methods. While we do not address optimization in this work, the formulation provides the necessary foundation for future studies of adaptive resource allocation in ETL systems.
\end{itemize}

\paragraph{\textbf{Outline of the article.}}
We begin in \Cref{sec:problem_setting} by formalizing the problem setting and introducing a graph representation of ETL process chains that serves as the foundation for a broader research program, starting with the present work. Next, in \Cref{sec:mathematical_model}, we introduce a detailed mathematical model of ETL processes starting in \Cref{sec:DES} with a mathematical framework for discrete-event simulation of ETL systems. This is followed by \Cref{sec:processing_times}, where we model the processing times based on the two central postulates \Cref{post:flow_balance} and \Cref{post: space of convenient curves} regarding the mean throughput and mean processing times, supported by empirical tests. We close both subsections by discussing directions for future research.

\paragraph{\textbf{Acknowledgments.}}
The authors thank Sven Behncke, Theresa Bick, Michael Brünker, Jan Langer, Giancarlo Meyer, Yvonne Reckmann, Henrik Riedel and  Nora Stroetzel for their ongoing support of this project and insightful discussions.
The authors acknowledge funding by the German Federal Ministry of Research, Technology and Space (BMFTR) under project ETL4Balance (FKZ 16IS23057A) and as part of the Robotics Institute Germany (RIG). Calculations for this research were conducted on the Lichtenberg high-performance computer of the TU Darmstadt.

\section{Problem Setting: ETL Process Chains}
\label{sec:problem_setting}
\begin{figure}
    \centering
    \includegraphics[width=0.6\textwidth]{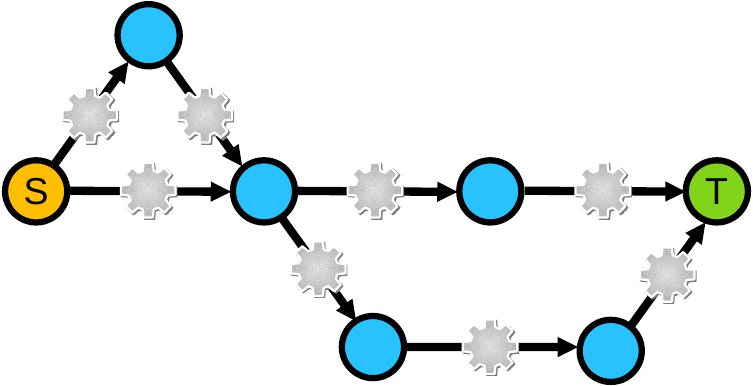}
    \caption{Directed acyclic graph $\mathcal{G} = (\mathcal{N}, \mathcal{E})$ representation of an ETL system. Circles represent data pools and gears represent ETL pipelines. This example has one overall source pool S (yellow node) and target pool T (green node), as well as multiple intermediate pools (blue nodes).}
    \label{fig:etl_graph}
\end{figure}

We consider process chains that iteratively process data records via different ETL pipelines. The processing includes sequential, interdependent steps and may involve alternative paths. For example, records containing either text or image data might go through mostly the same processing steps, but only images require an additional intermediate compression step. Such process chains form a directed acyclic graph (DAG)
\[
\mathcal{G} = (\mathcal{N}, \mathcal{E}),
\]
where nodes $u, v \in \mathcal{N}$ represent data storages or intermediate queues (data pools), and directed edges $e=(u,v) \in \mathcal{E}$ represent individual ETL pipelines connecting these components. Each edge $e=(u,v)$ represents an ETL operation consisting of three distinct phases: the data is extracted from the source pool $u$, transformed according to the edge's logic, and finally loaded into the target pool $v$. Each phase may use different amounts of computing resources, characterized by the number of threads available for processing multiple records at the same time. An example of this structure is illustrated in \Cref{fig:etl_graph}.  

\paragraph{Goal.}
Find a mathematical model of the flow of data records through such a process chain, suitable to describe and optimize the data throughput for minimal resource usage.

\section{Mathematical Modeling of ETL processes}
\label{sec:mathematical_model}
In this section, we describe the simulation framework used to model the throughput behavior of ETL pipelines. To this end, we employ a discrete-event simulation with a time resolution chosen sufficiently fine to resolve the shortest individual processing times involved, which are typically on the order of milliseconds. A detailed mathematical description of the simulation state and its time evolution under given resource allocations is provided in
\cref{sec:DES}.

Crucially, the model is not intended to reproduce all data processing details at the level of the underlying IT architecture. Instead, it is designed to capture the correct aggregate data throughput on time scales much larger than the simulation resolution—typically seconds to minutes. This is achieved by introducing a stochastic model for the time spent by a record in each individual phase of the ETL pipeline, which is described in \cref{sec:processing_times}.

\subsection{Discrete-Event Simulation}
\label{sec:DES}
Let $t \in \mathbb{N}$ denote the discrete time steps. At each time $t$, for given available computing resources, the current state of all data records in the the ETL process chain is sufficient to characterize how this state will be updated in the next time step, up to stochastic uncertainties in the processing times described in \cref{sec:processing_times}. That is, no knowledge of the past states is required. The ETL system can thus be modeled as a controlled Markov process (\cite{Puterman1994-rf})
\[
\bigl(\mathcal{S}, \mathcal{A}, \mathcal{P}\bigr),
\]
with state space $\mathcal{S}$, action space $\mathcal{A}$, and transition kernel $\mathcal{P}$.

\begin{definition}
The state at time $t$ is defined as
\[
s_t = \bigl(p_t^1, n_t^1, \ldots, p_t^N, n_t^N\bigr) \in \mathcal{S},
\]
where for each data record $i \in \{1,\dots,N\}$,
\[
p_t^i \in \{1,\dots,P\}
\quad\text{and}\quad
n_t^i \in \mathbb{N}_0
\]
denotes the current processing phase and the remaining processing time of record~$i$ in this phase, respectively. The $P$ different possible phases that a record can be in are the E, T and L phases of each individual ETL pipeline $e \in \mathcal{E}$, the different data pools $u \in \mathcal{N}$, and (depending on the specific ETL architecture) potentially also additional pipeline-internal queues between the E, T and L phases.
\end{definition}

\begin{definition}
The action at time $t$ is given by
\[
a_t = \bigl(a_t^{(1)}, \ldots, a_t^{(P_\mathrm{ETL})}\bigr) \in \mathcal{A},
\]
where $a_t^{(p)} \in \mathbb{N}$ denotes the number of parallel threads allocated to processing phase~$p$. This only includes the E, T and L phases, of which there are $P_\mathrm{ETL} = 3 \, |\mathcal{E}|$, as the remaining $P-P_\mathrm{ETL}$ phases only involve (temporarily) storing but not processing data.
\end{definition}

\begin{definition}
\label{def:transition}
The state-transition kernel factorizes over records and is given by
\[
\mathcal{P}\!\left(s_{t+1} \mid s_t, a_t\right)
= \prod_{i=1}^{N}
\mathcal{P}\!\left(p_{t+1}^i, n_{t+1}^i \mid s_t, a_t\right),
\]
with per-record transitions defined as
\[
p_{t+1}^i =
\begin{cases}
g_\mathcal{G}(p_t^i; s_t, a_t) & \text{if } n_t^i = 0,\\[0.3em]
p_t^i & \text{otherwise},
\end{cases}
\]
and
\[
n_{t+1}^i =
\begin{cases}
\max\{0,\, n_t^i - 1\}
& \text{if } p_{t+1}^i = p_t^i,\\[0.3em]
n_{t+1}^i \sim P_{p_{t+1}^i}\!\Bigl(\cdot \mid a_t^{(p_{t+1}^i)}\Bigr) & \text{otherwise}.
\end{cases}
\]
Here, $g_\mathcal{G}$ denotes the map from one phase to the next, according to the topology of the DAG $\mathcal{G}$, and $P_p$ is the phase-specific probability distribution of processing times per record, dependent on the new phase's resource allocation $a_t^{(p)}$ (cf.~\cref{sec:processing_times}).
\end{definition}

\Cref{def:transition} can be interpreted as follows. In each step, the remaining processing time steps in the current phase per record is reduced by 1 until it eventually hits 0. Once it reaches 0, the record moves to the next phase, as determined by the DAG-dependent map $g_\mathcal{G}$, which also takes into account the current state $s_t$ and actions $a_t$. The latter two are needed to determine if the record can actually move on to the next phase or not. If, for example, a record finished transforming (T), but there are no free load (L) threads (or no space in an optional queue between T and L), then the record must stay in the current T phase (and keep blocking a T thread) until space to move on frees up in a later time step. Furthermore, if a record reaches a final target pool, it will simply remain there. Once a record moves to a new phase, the total number of processing time steps in this new phase is sampled from the resource-dependent distribution $P_p$.

We close this section by noting that this Markovian discrete-event formulation with transitions factorizing over data records lends itself to an efficient implementation of this simulation using any high-performance computing language or library. A few considerations regarding simulation order have to be taken into account, though. To consistently track the progress of each record through the DAG, extract and load operations on the same data pool (node) may not be executed simultaneously. Moreover, since each pool serves as the source of all its outgoing edges and sink for its incoming edges, the order in which extract and load operations among these edges are executed is randomized at every simulation step, to prevent systematic biases in the form of preferably following one edge over the others. This randomization aligns more closely with the behavior of the real continuous-time system, where this is naturally caused by tiny stochastic variations of the per-record processing times.

\paragraph{Future directions.}
Given this setup, the flow of data records through an ETL process chain for any choice of actions $a_t$, i.e.~allocated number of parallel threads, can be simulated. In the simplest case those might be fixed over time or follow some predefined heuristic. In future work, we will investigate how a (probabilistic) policy function $\pi(a_t \mid s_t)$ can be learned that dynamically optimizes the allocated parallel resources to maximize throughput while minimizing resource usage.

\subsection{ETL Processing Times}
\label{sec:processing_times}
To complete the description of the discrete-event simulation in \cref{sec:DES}, we need a sufficiently accurate model of the processing time a data record spends in any of the three phases of a single ETL process. In practice, the processing time may depend on a multitude of factors, for example the current number of allocated parallel threads, the type and size of the record, the types of data stores involved, or environmental effects such as the background load of external requests reaching the data stores independently of the requests made by our ETL processes. Many of these effects are infeasible to model explicitly. So instead, we choose to stochastically model the processing times via parametrized probability distributions that can be heuristically calibrated to mimic the distribution of processing times observed in real ETL systems.

However, given our aim to use this simulation in the future for optimizing the resource allocation, the dependence of the processing times on the number of allocated parallel threads is one aspect that we want to model more explicitly.

To this end, we introduce queueing-theoretic concepts relevant to ETL workflows and postulate a Flow Balance law inspired by Little’s Law in mathematical queueing theory, introduced in~\cite{Little1961}. Little's Law provides a fundamental relationship between system occupancy, throughput, and processing time under strong assumptions such as stationarity and ergodicity. These assumptions are typically violated in ETL processes, and we do not claim any formal equivalence between Little’s Law and the model introduced here. Nevertheless, the high-level intuition behind Little’s Law---namely that parallelism, throughput, and processing time are intrinsically related---serves as a conceptual point of reference.

\begin{postulat}[{\textbf{Flow Balance}}]
\label{post:flow_balance} 
Consider any phase of an ETL process operating with $a$ parallel threads. We postulate that
\[
a = T(a)\cdot R(a) \qquad \forall \, a \in [0,\infty),
\]
where $T(a)$ denotes the mean throughput (number of processed data records per unit of time) and $R(a)$ denotes the mean processing time per record. \\
In particular, for all $a \in (0,\infty)$ with $T(a) > 0$, the mean processing time is given by
\[
R(a) = \frac{a}{T(a)}.
\] 
\end{postulat}

Next, motivated by real-world experience, we postulate a space of convenient curves to model the \emph{mean} throughput $T(a)$ of an ETL process operating with $a$ parallel threads. In practice, one expects the mean throughput to be monotonically increasing with growing $a$, starting from $T(0)=0$, but being bounded. That is, with more available threads the throughput generally increases, as more records can be processed in parallel, but not indefinitely. Due to non-parallelizable parts of the data processing or overhead distributing the work load over multiple threads, the maximal achievable throughput is limited. That is, for any curve $T$ modeling the mean throughput, there should exist
$T_{\max}$ such that
\[
\lim_{a \to \infty} T(a) = T_{\max} < \infty\, .
\]
Note that in most ETL processes, the throughput is usually already near $T_{\max}$ if $a$ is at the order of tens to hundreds of parallel threads.

With \Cref{post:flow_balance} in mind, the monotonicity requirement also ensures finite processing times when $a>0$. Although real-world systems may exhibit singular behavior in the throughput, e.g.~sudden drops to zero due to an overload of some IT architecture component when closely approaching $T_{\max}$, we deliberately exclude such cases from our modeling. Since our eventual objective is to optimize the throughput of an ETL process chain, such singular scenarios generally lie far above the range of relevant numbers of threads and are therefore avoided.

To this end, we postulate the following.

\begin{postulat}[\textbf{Space of Convenient Curves}]
\label{post: space of convenient curves}
Let the setting be as in \Cref{post:flow_balance}. Then the mean throughput $T$ is an element of the space of convenient curves $\mathcal{C}$, defined by
\[
\mathcal{C}
:= \left\{
f \in H^1_{\mathrm{loc}}\bigl([0,\infty),[0,\infty)\bigr)
\;\middle|\;
f' \ge 0 \text{ a.e. and } f \text{ is bounded}\, 
\right\}.
\] 
\end{postulat}

The $H^1_{\mathrm{loc}}$-regularity assumption in \Cref{post: space of convenient curves} provides a natural smoothness framework for the mean throughput function, ensuring sufficient regularity for gradient-based optimization, while excluding pathological functions, such as the Cantor function, which do not arise in real-world models.

For the reader’s convenience, we present natural families of curves that belong to the space considered in \Cref{post: space of convenient curves}.
\begin{example}
\label{ex: convenient curves}
    \item \textbf{Exponential family.}
    Curves of the form
    \[
    f : [0,\infty) \longrightarrow [0,\infty), \quad
    a \mapsto T_{\max}\left(1 - \mathrm{e}^{-a/k}\right),
    \qquad T_{\max}, k \in (0,\infty),
    \]
    belong to $\mathcal{C}$.

    \item \textbf{Rational curves.}
    Curves of the form
    \[
    f : [0,\infty) \longrightarrow [0,\infty), \quad
    a \mapsto \frac{x\, a}{y + a}, \qquad x,y > 0,
    \]
    or, more generally, any bounded rational curves on $[0,\infty)$ also belong to $\mathcal{C}$.
\end{example}

\begin{figure}[htbp]
    \centering
    \begin{subfigure}[t]{0.485\textwidth}
        \includegraphics[width=\textwidth,trim=0 -2ex 0 0]{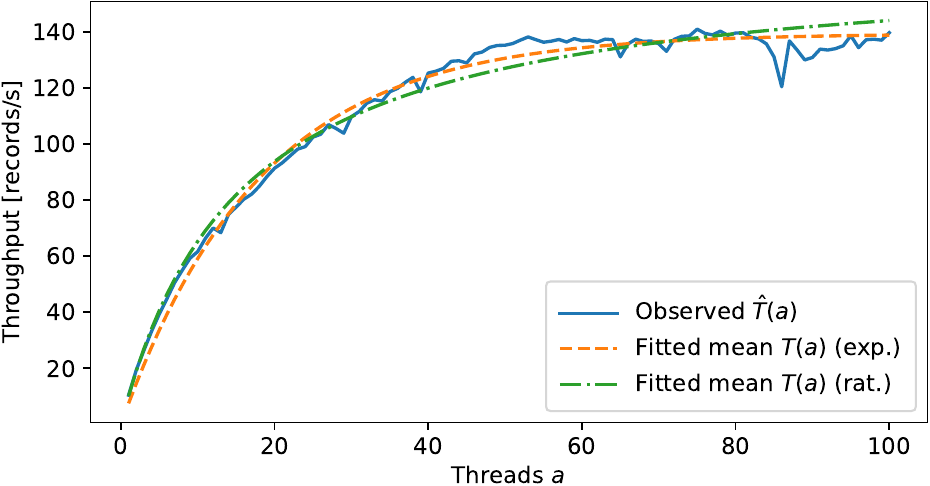}
        \caption{Observed throughput $\hat{T}(a)$ (blue) compared with a fit from the exponential family (orange) and a fit from the rational family (green), described in \Cref{ex: convenient curves}.\\}
        \label{fig:fitted T vs actual T}
    \end{subfigure}
    \hfill
    \begin{subfigure}[t]{0.485\textwidth}
        \includegraphics[width=\textwidth]{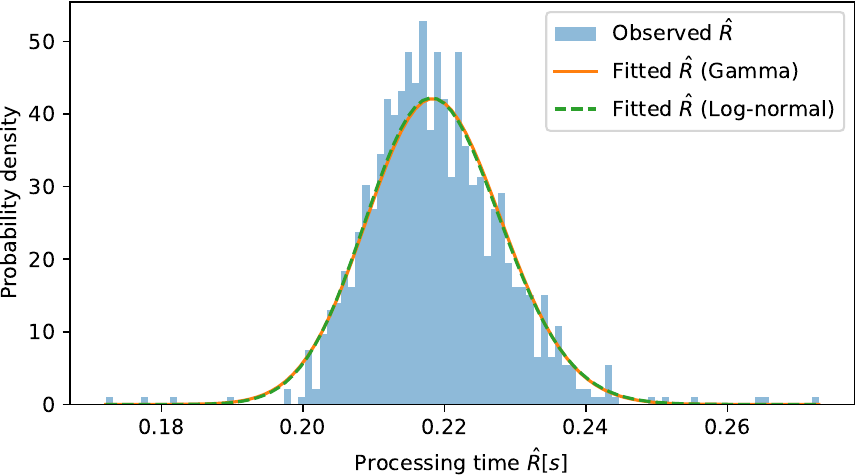}
        \caption{Observed processing times $\hat{R}$ for $a = 20$ threads (blue) compared with a fit of the gamma distribution (orange) and a fit of the log-normal distribution (green), described in \Cref{ex: processing time distrib}.}
        \label{fig:fitted R vs actual R}
    \end{subfigure}
    \vspace{-1\baselineskip}
    \caption{Observed data compared with fitted models.}
    \label{fig:fits_comparison}
\end{figure}

Before moving on, we provide in \Cref{fig:fitted T vs actual T} numerical evidence supporting \Cref{post:flow_balance} and \Cref{post: space of convenient curves} by explicitly testing them on the families of curves introduced in \Cref{ex: convenient curves}. In particular, \Cref{fig:fitted T vs actual T} compares observed throughput data $\hat{T}(a)$ with fitted models from the exponential and rational families. As can be seen, while the fitted curves capture the monotonic and bounded mean behavior predicted by the postulates, real-world observations $\hat{T}$ inevitably exhibit stochastic fluctuations around the idealized mean throughput $T$.

We therefore require a mathematical model for this stochastic noise. We begin by observing from real-world data that the processing time is inherently random. In particular, we require random variables that take only non-negative values, and empirical evidence suggests that the corresponding distributions often exhibit heavy tails. In the following, we provide examples of distributions that are promising candidates.
\begin{example}
\label{ex: processing time distrib}
\item \textbf{Gamma distribution.}
A natural candidate is the Gamma distribution,
\[
\hat{R}(a) \sim \Gamma\,\bigl(\alpha(a), \theta(a)\bigr),
\qquad a>0,
\]
with thread-number-dependent parameters $\alpha(a)>0$ and $\theta(a)=\frac{R(a)}{\alpha(a)}.$ The corresponding probability density function is
\[
f_{\hat R(a)}(x)=
\frac{1}{\Gamma\bigl(\alpha(a)\bigr)\,\theta(a)^{\alpha(a)}}\, 
x^{\alpha(a)-1} e^{-x/\theta(a)},
\qquad x>0.
\]

\item \textbf{Log-normal distribution.}
Another common candidate is the log-normal distribution,
\[
\hat{R}(a) \sim \mathrm{Lognormal}\,\bigl(\mu(a), \sigma^2(a)\bigr),
\qquad a>0,
\]
with thread-number-dependent parameters $\sigma(a)>0$ and $\mu(a)=\ln R(a)-\frac{\sigma(a)^2}{2}$. The corresponding probability density function is
\[
f_{\hat R(a)}(x)=
\frac{1}{x\,\sigma(a)\sqrt{2\pi}}
\exp\!\left(-\frac{\bigl(\ln x-\mu(a)\bigr)^2}{2\sigma^2(a)}\right),
\qquad x>0.
\]
\\
For both distributions, the parametrization is chosen such that the expectation value of the processing time matches the desired mean value,
\[
\mathbb{E}\!\left[\hat{R}(a)\right] = R(a) \qquad \forall a \in (0,\infty).
\]
\end{example}

In \Cref{fig:fitted R vs actual R} we present tentative numerical evidence for these types of distributions. Based on the same observed data used in \Cref{fig:fitted T vs actual T}, we plot the histogram of observed processing times $\hat{R}$ for $a=20$ threads. Fitting the distributions in \Cref{ex: processing time distrib} to these data yields nearly identical results and provides a qualitatively good match.

To obtain the number of processing time steps $n_{t+1}$ in the transition kernel of our discrete-event simulations defined in \Cref{def:transition}, the continuous processing time samples $\hat{R}$ simply are rounded to the nearest whole number of simulated time steps.

We conclude this subsection by noting that the extract-phase processing times in an ETL process may behave differently than just described, as often this phase is implemented using batched extraction of multiple records at once, just using a single thread, as well as caching of previously extracted records to accelerate the extraction. In such accelerated setups the extract time often is considerably smaller than the transform and load times and thus only plays a minor role for the overall throughput. Thus, modeling the extract times with a simplified thread-number-independent, uniform distribution often suffices,
\[
\hat{R}_\mathrm{E} \sim \mathcal{U}\left\{R_{\min}, R_{\max}\right\},
\]
with lower and upper bounds $R_{\min}, R_{\max}$.

\paragraph{Future directions.}
Despite support for \Cref{ex: processing time distrib} in first empirical tests, a more systematic treatment of stochastic modeling choices remains an interesting direction for future work. In particular, it would be desirable to postulate a framework similar in spirit to \Cref{post:flow_balance} and \Cref{post: space of convenient curves}. A natural first step would be to collect qualitative and quantitative properties from a wider range of observed processing-time distributions and then derive a specific family of suitable probabilistic models based on these properties.

Additionally, we plan to further investigate more realistic modeling choices for extract phases that involve batched extraction or caching. Although we expect this to only have a minor impact on the optimization of many ETL processes, it becomes more relevant in cases where extract times contribute notably to the total processing time.

\bibliographystyle{gtml2025_workshop}
\bibliography{gtml2025_workshop}

\end{document}